\documentclass[aps,prl,twocolumn,preprintnumbers,amsmath,amssymb]{revtex4-1}

\usepackage{graphicx}
\usepackage{bm}
\usepackage{upgreek}

\begin{document}


\title{Superconducting condensate residing on small Fermi pockets in underdoped cuprates}

\author{V. M. Krasnov }
\email[E-mail: ]{Vladimir.Krasnov@fysik.su.se}


\affiliation{Department of Physics, Stockholm
University, AlbaNova University Center, SE-10691 Stockholm,
Sweden}

\begin{abstract}
How does Fermi surface develop in cuprates upon doping of a parent
Mott insulator, does it consist of large barrels or small pockets,
which of them is responsible for superconductivity and what is a
role of the pseudogap? Those are actively debated questions,
important for understanding of high temperature superconductivity.
Here we analyze doping dependence of interlayer tunneling in
cuprates. We
observe that with decreasing doping the supercurrent is rapidly
decreasing, but the quasiparticle resistance at high bias remains
almost unchanged. This indicates that Cooper pair and
quasiparticle currents originate from different parts of Brillouin
zone: Cooper pairs are residing only on small pockets, which are
progressively shrinking with decreasing doping, but the
quasiparticle current is integrated over the full length of
barrels, which are only weakly doping dependent. The expanding
pseudogap areas along the barrels do not contribute to pair
current. This provides direct evidence for nonsuperconducting
origin of the pseudogap.

\end{abstract}



\maketitle

Fermi surface in metals occurs at an intersection of a conduction
band with a chemical potential. Insulators do not have Fermi
surface because the chemical potential lies in the band gap
region. High temperature superconductivity in cuprates appears
upon doping of a Mott insulator. One of the key questions is how
Fermi surface develops with doping
\cite{Pickett_1992,Damasselli_2003}.
According to Luttinger theorem 
Fermi surface area should be proportional to doping $p$. However,
for cuprates the photoemission edge, seen by angular-resolved
photoemission spectroscopy (ARPES), does not grow gradually with
doping but forms large barrels with an area $\propto (1+p)$
already at the lowest doping
\cite{Damasselli_2003,Kordyuk_2002,Ronning_2003}.
However, strictly speaking only nodal parts of the barrels (Fermi
arcs) are representing true Fermi surface because anti-nodal parts
are gapped \cite{Ronning_2003,Kanigel_2006}. The arcs size is
growing linearly with doping
\cite{Lee_2007,Ideta_2012,Vishik_2012,Kohsaka_2008}. Luttinger
theorem can be satisfied assuming that arcs represent parts of
small Fermi pockets \cite{Ronning_2003,Meng_2009,Yang_2011}.
Although existence of small pockets was confirmed by quantum
oscillation experiments
\cite{Vignolle_2011,Yelland_2008,Riggs_2011,Barisic_2013}, their
position in Brillouin zone \cite{Pickett_1992}, connection to
barrels and significance for high-$T_c$ superconductivity remains
unclear \cite{Chakravarty_2011}.

The role of a normal state pseudogap (PG) is another related issue
\cite{TallonPhC,Norman_2005}. Similarities between the PG and the
superconducting gap (SG) have led to an assumption of precursor
superconductivity origin of the PG \cite{Valla_2008}. In this case
anti-nodal PG parts of barrels should contain a major part of the
superconducting condensate. However, there are also arguments in
favor of competition of the two co-existing gaps
\cite{KrTemp,SecondOrder,Vishik_2012,MR,Jacobs_Bi2201,TallonPhC,Doping,Millis_2013,Gabovich,Chakravarty_2011}.
Discrimination between SG and PG is particularly difficult at
$T<T_c$, when the whole barrel
is gapped. So far it is not possible to conclude whether
superconductivity is originating from large barrels or small Fermi
pockets.

Here we analyze doping dependence of interlayer tunneling
characteristics of small
Bi$_2$Sr$_2$Ca$_{1-x}$Y$_x$Cu$_2$O$_{8+\delta}$ [Bi(Y)-2212]
intrinsic Josephson junctions. We utilize the ability of
superconducting tunnel junctions to independently probe Cooper
pair and single quasiparticle (QP) currents. We observe that the
high-bias QP resistance remains almost doping independent,
implying that QP current is originating from weakly doping
dependent barrels. To the contrary, the supercurrent is rapidly
decreasing with decreasing doping, indicating that Cooper pairing
occurs only on small Fermi pockets that progressively shrink with
decreasing doping. The antinodal parts of barrels, which grow with
underdoping, do not contribute to supercurrent. This directly
proves that the PG is not due to Cooper pairing. We present
numerical calculations
that support our conclusions.

\begin{figure*}[t]
    \centering
    \includegraphics[width=0.9\textwidth]{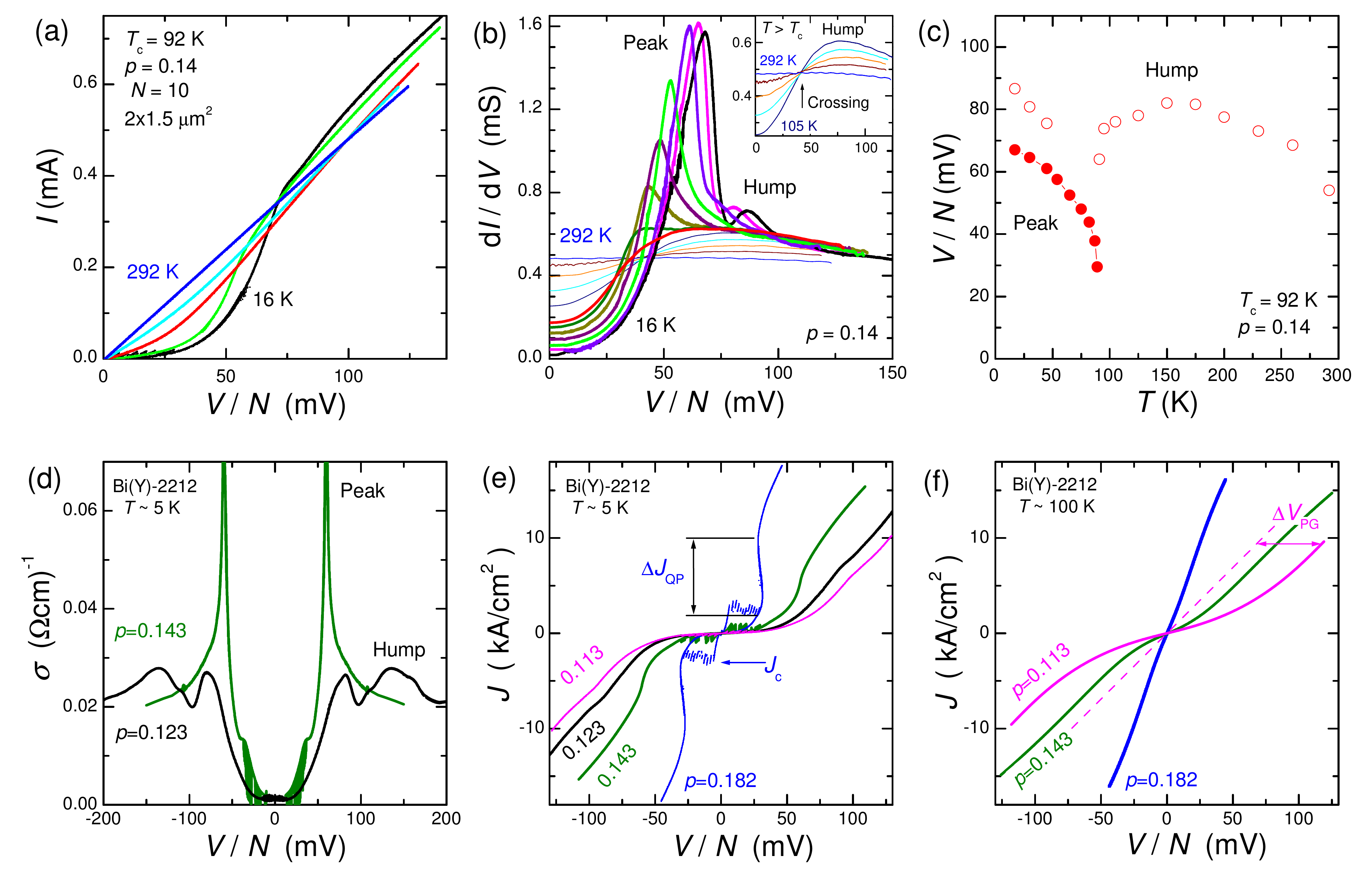}
    \caption{(Color online) Temperature and doping dependence of intrinsic tunneling characteristics.
(a) $I$-$V$ and (b) $dI/dV(V)$ characteristics of a small
moderately underdoped mesa at different $T$. Inset demonstrates
crossing of $dI/dV(V)$ curves at $T>T_c$. (c) Temperature
dependencies of the superconducting peak and the pseudogap hump
voltages for the same mesa. (d) Comparison of normalized intrinsic
spectra for slightly and strongly underdopoed mesas. (e) and (f)
Comparison of normalized $I$-$V$ curves at different doping (e) in
the superconducting and (f) in the normal states.}
    \label{fig:fig1}
\end{figure*}

Intrinsic Josephson junctions are naturally formed in Bi-2212
single crystals \cite{Kleiner}.
Atomic scale of such junctions leads to a large capacitance
and a quality factor $Q \sim 10^2$ \cite{Katterwe_2010}.
Current-Voltage ($I$-$V$) characteristics of junctions with $Q \gg
1$ acquire a hysteresis with zero-voltage and resistive branches
corresponding to Cooper pair and QP tunneling, respectively,
\cite{KrTemp,SecondOrder,MR,Suzuki_2013,Irie_2002,Inomata_2003,Ren_2012,Katterwe_PRL2008}.
which allows independent analysis of pair and QP transport.

We study Bi(Y)-2212 single crystals from the same batch with an
optimal $T_c(OP) \simeq 95$ K. Doping level $p$ was changed by
annealing at $T=600^{\circ}$C and was estimated using an empirical
expression $T_c(p)=T_c(OP)[ 1-82.6(p-0.16)^2 ]$ \cite{TallonPhC}.
It corresponds to an onset of superconductivity at $p=0.05$ at the
insulator-to-metal transition and to the optimal doping (OP) at
$p=0.16$ (holes per Cu). Micron-size mesa structures, containing
$N=7-12$ junctions were made
using micro/nano-fabrication techniques. Details of sample
fabrication and characterization can be found in Refs.
\cite{KrTemp,SecondOrder,MR,Katterwe_PRL2008}.

\begin{figure*}[t]
    \centering
    \includegraphics[width=0.98\textwidth]{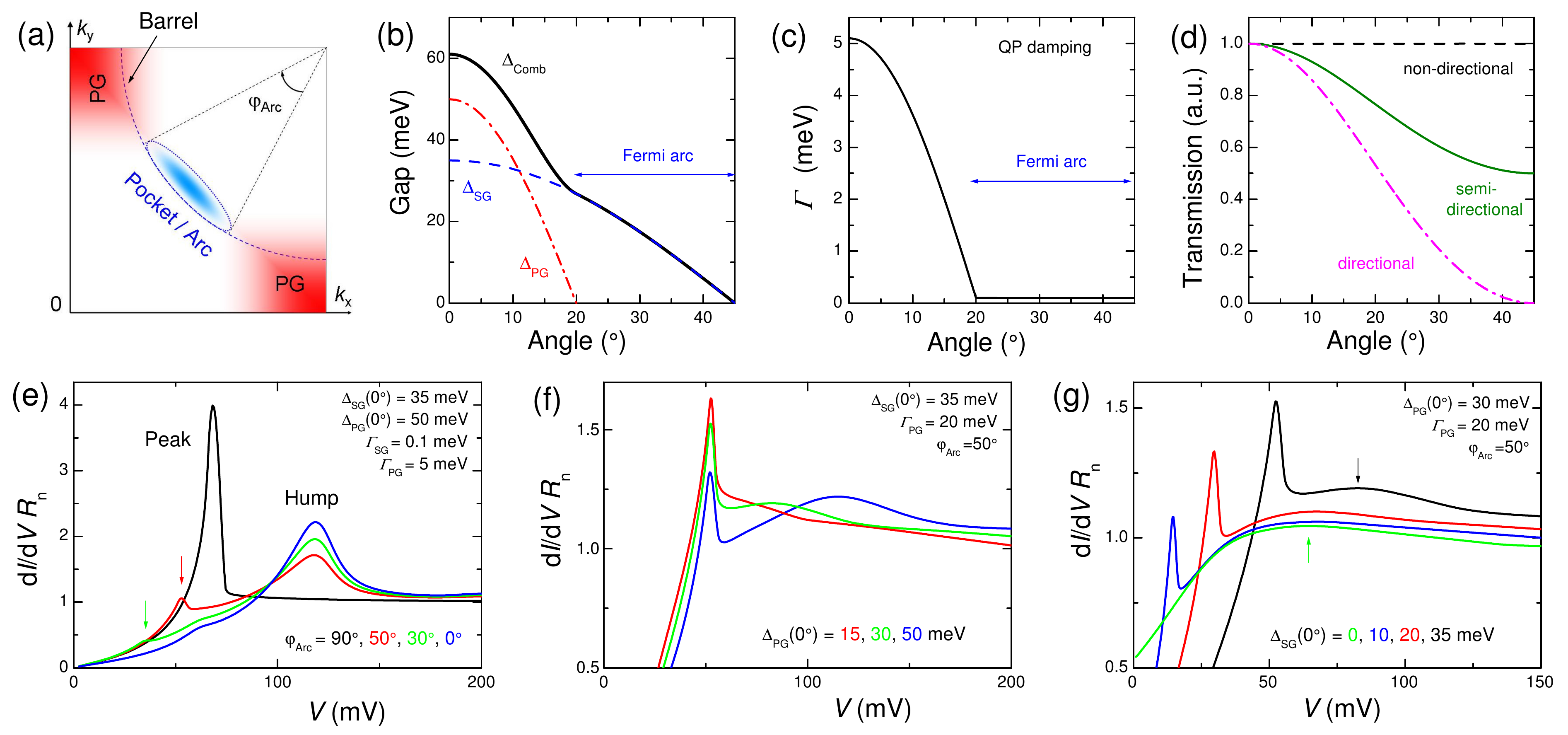}
    \caption{(Color online) Numerical modelling of intrinsic tunneling characteristics.
(a) Sketch of a quarter of the Brillouin zone. (b)-(d) Angular
distribution along the barrel from an antinodal to a nodal point
of (b) the gaps, (c) the quasiparticle damping factor and (c) the
tunneling transmission probability. (e)-(g) Evolution of
calculated $dI/dV(V)$ curves upon varying of (e) the pocket size,
(f) The pseudogap energy and (g) the superconducting gap. Arrows
indicate positions of peaks and humps.}
    \label{fig:fig2}
\end{figure*}

Figure \ref{fig:fig1} summarizes temperature and doping
dependencies of interlayer tunneling. Fig. \ref{fig:fig1} (a)
shows $T$-dependencies of $I$-$V$'s for a moderately underdoped
(UD) mesa $p\simeq 0.14$.
At $T<T_c$ a sum-gap kink appears at $V=2\Delta/e$ (per junction)
followed by Ohmic, and almost $T$-independent tunnel resistance
$R_n$. The constancy of $R_n$ is a fundamental consequence of
electronic state conservation \cite{MR}. Spectroscopic features
are better analyzed using $dI/dV(V)$ characteristics, shown in
Fig. \ref{fig:fig1} (b). It is seen that the sum-gap peak shifts
to lower voltages and decreases in amplitude with increasing $T$
and vanishes at $T_c$. At $T>T_c$ the zero-bias conductance
remains suppressed as a consequence of persisting PG. The
corresponding missing states are expelled into a broad hump at
higher voltages. With increasing $T$ the zero bias minimum
fills-in and the hump is deflated in a state conserving manner
\cite{SecondOrder,MR}, so that the $dI/dV(V)$ characteristics at
different $T$ cross at one point, as shown in the inset
\cite{Katterwe_PRL2008,SecondOrder}.
Fig. \ref{fig:fig1} (c) shows $T$-dependencies of the
superconducting peak and the pseudogap hump. It is seen that the
SG decreases rapidly upon approaching the $T_c$ in a BCS
(Bardeen-Cooper-Schrieffer) manner \cite{SecondOrder}. The PG
persists in a broad $T$ range both below and above $T_c$.

Fig. \ref{fig:fig1} (d) shows differential conductivity
for a slightly $p=0.143$ and strongly
$p=0.123$ UD mesas at low $T$.
It is seen that for a strongly UD mesa $p=0.123$  the hump
coexists with the peak at $T\ll T_c$ \cite{KrTemp,Doping}, the
peak is strongly suppressed, the hump is enhanced and has even
larger amplitude than the peak. With increasing doping the peak
height is increasing and the hump is decreasing both in height and
voltage. 
With further increase of doping the hump is buried under the peak,
as seen from Fig. \ref{fig:fig1} (d) for $p=0.143$. The hump,
however, is uncovered at elevated $T$ as the peak shifts to lower
voltages \cite{KrTemp}.

\begin{figure*}[t]
    \centering
    \includegraphics[width=0.9\textwidth]{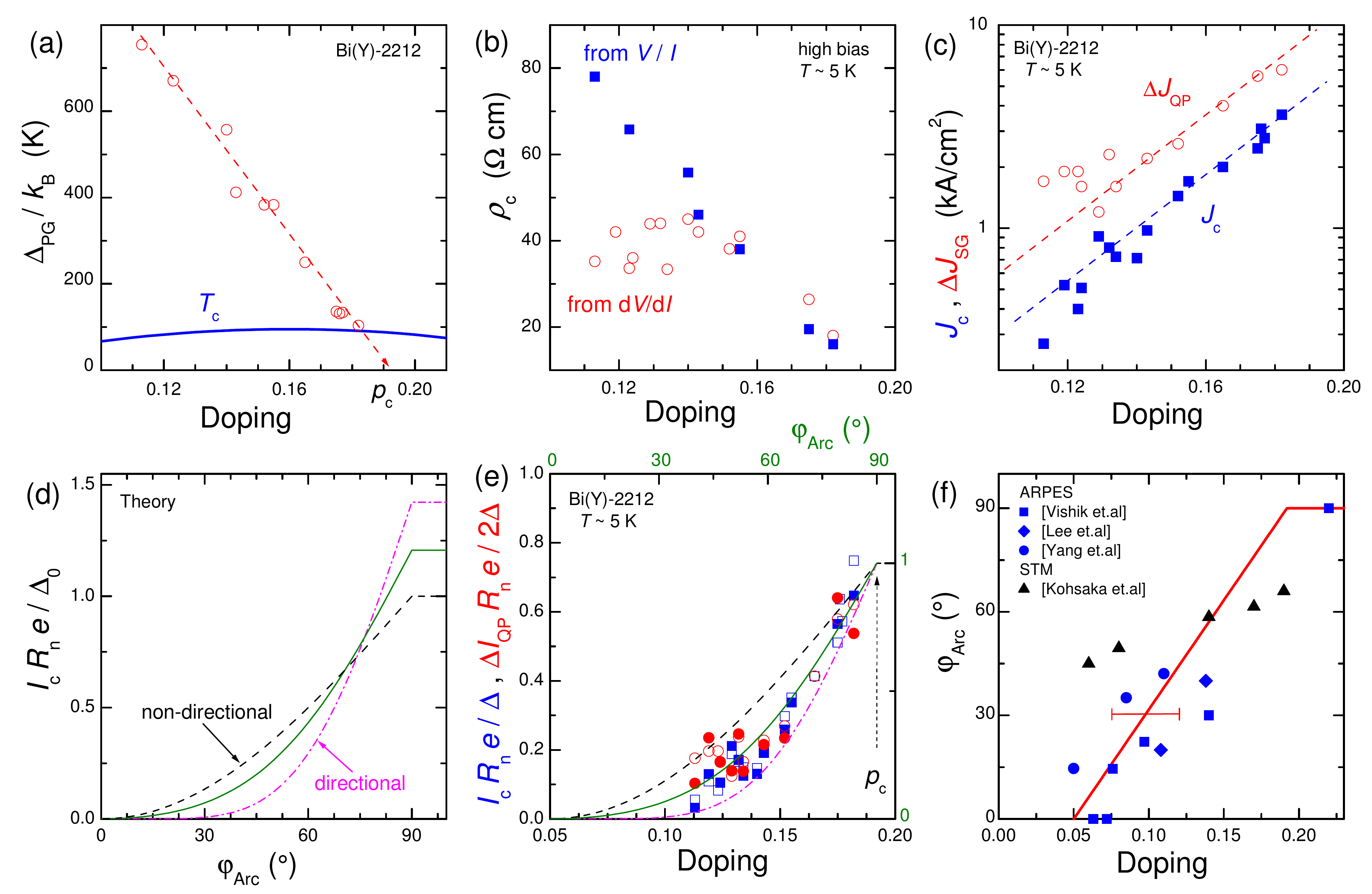}
    \caption{(Color online) Doping dependence of (a) the pseudogap, (b) high-bias
resistivities, (c) the critical current density $J_c$ and the
amplitude of the sum-gap kink $\Delta J_{QP}$. (d) Calculated
dependence of the critical current from Fermi pockets, as a
function of $\varphi_{Arc}$ for non-directional, semi-directional
and directional tunneling (dashed, solid and dashed-dotted lines).
(e) Comparison of scaled $I_c$ and $\Delta I_{QP}/2$. They
represent the amounts of Cooper pairs and QP states subjected to
pairing, respectively.
Lines (top-right axes) represent normalized curves from panel (d).
The reduction of $I_c$ and $\Delta I_{QP}$ with decreasing doping
is consistent with proportional shrinkage of the Fermi pockets
containing the superconducting condensate. (f) The solid line
represents a deduced doping dependence of the pocket size.
Symbols represent corresponding data obtained by other techniques.
}
    \label{fig:fig3}
\end{figure*}

Fig. \ref{fig:fig1} (e) and (f) represent current density $J$ vs.
voltage per junction characteristics for different doping levels
(e) at low $T$ and (f) above $T_c$.
The following main features, which will be in focus of our
discussion, are seen:

(i) {\em The critical current.} Multiple branches at low bias
appear due to one-by-one switching of junctions from the
superconducting to the resistive state \cite{Kleiner,KrTemp}. The
amplitude of the branches represent the Josephson critical current
density $J_c$, marked in Fig. \ref{fig:fig1} (e). $J_c$ rapidly
decreases with decreasing doping. As a result, branches, which are
very pronounced for the overdoped (OD) mesa $p=0.182$, are hardly
visible (on this scale) for the most UD $p=0.113$ mesa. Reduction
of $J_c$ reflects reduction of the superconducting condensate with
decreasing doping.

(ii) {\em The sum-gap kink amplitude}. $\Delta J_{QP}$ represents
a number of QP states within the superconducting gap, which are
subjected to pairing. 
Therefore,
$\Delta J_{QP}$ and $J_c$ should be directly connected.
Indeed, from Fig. \ref{fig:fig1} (e) it is seen that $\Delta
J_{QP}$ is also rapidly decreasing with decreasing doping so that
the kink becomes poorly visible for the most UD mesa. This
indicates a reduction of the number of electronic states subjected
to Cooper pairing. 

(iii) {\em The pseudogap} is most clearly visible in the normal
state when superconducting features are gone. From Fig.
\ref{fig:fig1} (f) it is seen that the $I$-$V$ of a strongly UD
mesa $p=0.113$ is nonlinear at $T>T_c$. The current is suppressed
below a threshold voltage $\Delta V_{PG}$.
$\Delta V_{PG}$ decreases with increasing
doping and disappears for slightly OD mesas, as seen from an
almost Ohmic $I$-$V$ at $p=0.182$.

(iv) {\em The tunnel resistance.} From Fig. \ref{fig:fig1} (e) it
is seen that the high-bias $R_n$ is initially increasing with
decreasing doping, but for UD mesas the differential resistance
becomes almost doping independent.
Since tunnel resistance depends on the
QP DoS it is expected that $R_n$ should increase with decreasing
carrier concentration and saturation at low doping requires
explanation.

To gain a better understanding of experimental characteristics we
performed numerical analysis 
taking into account a 
symmetry of the order parameter and a topology of the Fermi
surface, as shown in Fig. \ref{fig:fig2} (a).
Following recent studies \cite{Meng_2009,Yang_2011,Vignolle_2011},
we assume that central parts of large barrels represent small
Fermi pockets with an angular size $\varphi_{Arc}$. Fig.
\ref{fig:fig2} (b) represents angular dependencies of the gaps
along the barrel. In the normal state only antinodal PG parts of
barrels are gapped. 
At $T<T_c$ a d-wave SG
is opening. 
We assume that the antinodal regions acquire a combined gap
$\Delta_{Comb}=\sqrt{\Delta_{PG}^2+\Delta_{SG}^2}$.
The exact scenario in the antinodal region is not very important
because the main difference between barrel and arc regions is in
the QP damping factor $\Gamma$, which is small at the pocket and
large at the barrel, as shown in Fig. \ref{fig:fig2} (c). This
makes QP's ill-defined in the PG region \cite{Damasselli_2003}.
Finally, we have to take into account angular dependence of the
transmission probability $Tr(\varphi)$. For non-directional
tunneling $Tr(\varphi)=const$. However, $c$-axis transport in
cuprates should be dominated by antinodal regions. 
The corresponding directional $Tr(\varphi)$ is
shown by the dashed-dotted line in Fig. \ref{fig:fig2} (d). We
have found that the experimental data is best fitted using a
semi-directional $Tr(\varphi)$, which is finite at nodal regions,
as shown by the solid line in Fig. \ref{fig:fig2} (d). Numerical
simulations presented in Figs. \ref{fig:fig2} (e-g) are made for
the semi-directional case. Details of calculations can be found in
the Supplementary \cite{Supplem}.

Fig. \ref{fig:fig2} (e) demonstrates variation of $dI/dV(V)$ upon
changing of the arc size. 
The case $\varphi_{Arc}=90^{\circ}$
corresponds to the absence of the PG. 
In this case the spectrum contains a single superconducting peak
at $eV=2\Delta_{SG}(0^{\circ})$ with the height and the shape
similar to that for OP mesas.
As the arc shrinks, the amplitude of
the peak is rapidly decreasing and the PG hump is growing at
$eV=2\Delta_{Comb}(0^{\circ})$. This is similar to evolution of
experimental curves with decreasing doping in Fig. \ref{fig:fig1}
(d).

Fig. \ref{fig:fig2} (f) demonstrates variation of $dI/dV(V)$ upon
changing the PG energy. When $\Delta_{PG}$ is significantly larger
than $\Delta_{SG}$ both the peak and the hump are well
defined. However, as $\Delta_{PG}$ starts to decrease 
the hump is moving towards the peak and is eventually buried under
the peak. This is similar to evolution of experimental curves with
increasing doping, see Figs. \ref{fig:fig1} (b) and (d).

Fig. \ref{fig:fig2} (g) shows variation of $dI/dV(V)$ upon
changing of $\Delta_{SG}$, mimicking $T$-variation shown in Fig.
\ref{fig:fig1} (b). Note that the hump moves to lower voltages
with decreasing $\Delta_{SG}$ because in our case it occurs at the
combined gap. 
We observe a similar shift of the hump at $T<T_c$ in experiment,
as shown in Fig. \ref{fig:fig1} (c). We conclude that there is a
good overall agreement between the experimental data and the
considered model.
Therefore, we employ this model for a quantitative analysis of
data.

Figure \ref{fig:fig3} represents a summary of doping dependence of
intrinsic tunneling characteristics. Fig. \ref{fig:fig3} (a) shows
the PG energy obtained from half of the PG hump voltage. It is
linearly decreasing with increasing doping and tends to vanish at
the quantum critical point slightly above $p_c=0.19$, in agreement
with Refs. \cite{TallonPhC,Vishik_2012}. A topological
barrel-pocket transition should occur at this point
\cite{Vignolle_2011}.

In Fig. \ref{fig:fig3} (b) we analyze doping dependence of high
bias resistivity. Solid and open symbols represent dc and
differential (ac) values, respectively.
From Fig. \ref{fig:fig1} (f) it is seen that for OD mesas the
$I$-$V$ is Ohmic and the two resistances coincide. The $I$-$V$'s
of UD mesas become non-linear 
due to appearance of the PG and the two resistances become
different. The linear doping dependence of the dc resistivity
reflects the corresponding behavior of the PG, Fig. \ref{fig:fig3}
(a). However, the ac resistivity becomes almost doping independent
at low doping, as seen from Fig. \ref{fig:fig1} (e).
Simulations provide a clarification of different behavior of dc
and ac resistivities. From Figs. \ref{fig:fig2} (e) and (f) it is
seen that the high-bias $dI/dV$ is independent of $\varphi_{Arc}$
and $\Delta_{PG}$. This occurs because the PG does not change the
total amount of states but just redistributes them. At a
sufficiently high voltage $eV\gg \Delta_{PG}$ all QP states along
the barrels, including the gapped anti-nodal parts, contribute to
the QP current. Therefore, the weak doping dependence of
ac-resistivity is a consequence of the weak doping
dependence 
of barrels \cite{Ronning_2003,Kordyuk_2002}.

Fig. \ref{fig:fig3} (c) shows doping dependence of $J_c$ and
$\Delta J_{QP}$
It is seen that unlike $R_n$ both $J_c$ and $\Delta J_{QP}$ decay
rapidly with decreasing doping. Behavior of $J_c(p)$ is in
agreement with previous reports
\cite{Doping,Irie_2002,Suzuki_2013,Inomata_2003}. A qualitative
difference of doping dependencies of the Cooper pair and the
high-bias QP transport is the main new observation of this work.
As explained above, the weak doping dependence of $R_n$
suggests that the QP current is integrated over the full length of
barrels, which are weakly doping dependent. Therefore, a rapid
decrease of the Cooper pair current with decreasing doping
indicates that the superconducting condensate does not reside
along the full length of barrels, but occupy a progressively
smaller fraction, as expected for small Fermi pockets.

To understand how $I_c$ depends on the pocket size we performed
corresponding calculations, shown in Fig. \ref{fig:fig3} (d)
(see Ref. \cite{Supplem}). In the non-directional case,
\begin{equation}\label{IcRn}
I_c =\frac{\Delta_0}{e R_n}\left[1-\cos(\varphi_{Arc}) \right].
\end{equation}
In the absence of the PG, $\varphi_{Arc}=90 ^{\circ}$,
$I_c R_n=\Delta_0/e$. With decreasing doping Fermi arcs shrink and
$I_c$ decreases. The $I_c$ vanishes when arcs collapse
$\varphi_{Arc}\rightarrow 0$.
The supercurrent and the high-bias QP resistance have different
doping dependencies because supercurrent is measured at zero
voltage. Consequently, pair tunneling occurs only between true
(ungapped) Fermi surfaces, i.e., Fermi pockets, which shrink with
decreasing doping. To the contrary, QP tunneling occurs at finite
bias and at $eV\gg \Delta_{PG}$ it collects all QP states along
the barrels.

From Fig. \ref{fig:fig3} (c) it is seen that $\Delta I_{QP}$ and
$I_c$ are changing in a correlated manner. For a d-wave tunneling
in the absence of the PG the kink amplitude should be twice the
critical current $\Delta I_{QP} =2\Delta_0/e R_n = 2 I_c$. In Fig.
\ref{fig:fig3} (e) we show the correspondingly scaled quantities
as a function of doping. They were obtained using the high bias
differential resistance as $R_n$ (open and solid symbols were
obtained using different criteria for estimation of $R_n$ and
$\Delta$). It is seen that $I_c$ and $\Delta I_{QP}/2$ merge
together, confirming that both are indeed related: $I_c$ is
measuring the amount of Cooper pairs and $\Delta I_{QP}$ the
amount of QP states within the gap, subjected to pairing.
Importantly, the scaling indicates that both quantities originate
from the same parts of the Brillouin zone ( i.e., Fermi pockets).
The reduction of both $I_c$ and $\Delta I_{QP}$ with decreasing
doping reflects the shrinkage of the pocket size.

In order to estimate variation of $\varphi_{Arc}$ with doping, in
Fig. \ref{fig:fig3} (e) we compare experimental data with
theoretical curves from Fig. \ref{fig:fig3} (d).
Here we assumed that $\varphi_{Arc}=90^{\circ}$ at the onset of
the PG, $p = 0.192$, and that $\varphi_{Arc}=0^{\circ}$ at the
insulator-to-metal transition, $p=0.05$. The agreement is
remarkable taking into account that there is no fitting other than
adjustment of the vertical scale.

Figure \ref{fig:fig3} (f) summarizes our main result. A solid line
represents the deduced angular size of the Fermi pocket/arc as a
function of doping. It shrinks linearly with decreasing doping,
consistent with previous reports by other techniques
\cite{Vishik_2012,Lee_2007,Yang_2011,Kohsaka_2008}, shown for
comparison in the figure. An important new aspect of our work is
that we could discriminate between Cooper pair and quasiparticle
transport. We conclude that the QP current at high bias is
originating from the full area of large barrels, leading to weak
doping dependence of $R_n$. However, the Cooper pair current
originates only from small Fermi pockets, which shrink with
decreasing doping, leading to a rapid decrease of both the
critical current and the sum-gap kink amplitude. The pseudogap
parts of barrels, which grow bigger with underdoping, apparently
do not contribute to supercurrent. This is a direct evidence for
non-superconducting origin of the pseudogap in cuprates.

Technical support from the Core Facility in Nanotechnology at SU
is gratefully acknowledged.

\section{Supplementary Information}

The supplementary material provides additional information about
numerical calculations of interlayer tunneling characteristics and
analytical results for variation of $I_c R_n$ as a function of the
arc size for non-directional, directional and semi-directional
energy and momentum conserving tunneling between d-wave
superconductors. We assumed that the supercurrent is originating
only from nodal arcs and single quasiparticle current from the
full length of large barrels.

\subsection{Calculation of interlayer tunneling characteristics}

Proper calculation of interlayer tunneling characteristics
requires accurate integration of tunneling current over all
initial and final states, taking into account the band structure,
angular dependence of the energy gap $\Delta(\varphi)$ and the
transmission probability $Tr(\varphi_1,\varphi_2)$ for tunneling
between the initial and final states with the momentum angles
$\varphi_{1,2}$, where $\varphi$ is the angle between the momentum
of the QP and the principle axis of the Brillouin zone. We
consider the case of elastic tunneling, in which the energy of
electrons is conserved. In this case the QP tunneling current
between two adjacent superconducting layers can be written as:
\begin{eqnarray}\label{S1Eq.Tunn}
I = A \int_0^{2\pi}\frac{d\varphi_1}{2\pi} \int_0^{2\pi} \frac{d\varphi_2}{2\pi} \int_{-\infty}^{+\infty} dE~~~~~~~~~ ~ \\
\nonumber
Tr^2(\varphi_1,\varphi_2)N(E,\varphi_1)N(E+eV,\varphi_2)\left[f(E)-f(E+eV)
\right],
\end{eqnarray}
where $E$ is the energy of the QP with respect to the chemical
potential, $N(E,\varphi_1)$ and $N(E+eV,\varphi_2)$ are the
corresponding QP DoS in the initial and final states:
\begin{equation}\label{S2N(G)}
N(E,\varphi)=N(0)\Re\left[\frac{E-i\Gamma(\varphi)}{\sqrt{(E-i\Gamma(\varphi))^2-\Delta(\varphi)^2}}\right].
\end{equation}
Here $\Gamma(\varphi)$ is the angular dependent quasiparticle
damping factor (the inverse QP lifetime). In the absence of the
PG, analysis of various tunneling scenarios for coherent
$\varphi_1=\varphi_2$ and incoherent $\varphi_1 \ne \varphi_2$
tunneling between d-wave superconductors can be found in the
supplementary material to Ref. \cite{Katterwe_PRL2008} and in Ref.
\cite{Yamada}. Here we will focus on analysis of the pseudogap
effect for the case of coherent $\varphi_1=\varphi_2$ and elastic
(momentum and energy conserving) tunneling for a model of a
``remnant" Fermi barrel with a PG in antinodal and a Fermi-arc in
the nodal parts, as shown in Fig. 2 (a). We assume that Cooper
pair tunneling occurs only in the Fermi-arc regions, but the QP
tunneling occurs along the whole barrel. All the characteristics
are presented for a quarter of Brulloin zone and are symmetrically
reflected for other three quarters of the zone.

Fig. 2 (b) represents angular dependence of the gaps. In the
normal state only PG is present in the antinodal parts of the
barrels in the angular interval from 0 to
$\varphi_{PG}=\pi/4-\varphi_{Arc}/2$. We assumed the following
angular dependence,
\begin{equation}\label{S3D_PG}
    \Delta_{PG}(\varphi)=\Delta_{PG}(0)\cos\left(\frac{\pi
    \varphi}{2\varphi_{PG}}\right),
\end{equation}
shown by the dashed-dotted line in Fig. 2 (b).

At $T<T_c$ there is a superconducting gap with a d-wave symmetry,
\begin{equation}\label{S4d_wave}
\Delta_{SG}(\varphi)=\Delta_0 \cos(2\varphi),
\end{equation}
shown by the dashed line in Fig. 2 (b). We assume that in the PG
region the two gaps form a combined gap
\begin{equation}\label{S5Dcomb}
\Delta_{Comb}(\varphi)
=\sqrt{\Delta_{PG}^2(\varphi)+\Delta_{SG}^2(\varphi)},
\end{equation}
shown by the solid line in Fig. 2 (b). This assumption is,
however, not critical because antinodal parts of the barrel do not
contribute to a sharp sum-gap peak in $dI/dV$ because of a large
QP damping in the PG region \cite{Damasselli_2003}, as shown in
Fig. 2 (c). We assumed that within the PG region the $\Gamma_{PG}$
is varying in a similar manner to Eq. (\ref{S3D_PG}) with the
maximum $\Gamma_{PG}(0) \sim 10$meV. In the arc region the QP
damping is two orders of magnitude smaller $\Gamma_{SG}=0.1$meV.
Due to the large difference in $\Gamma$ sharp sum-gap kink/peak
features in $I$-$V$ and $dI/dV$ originate solely from the arc
region and correspond to the superconducting gap. Eq.
(\ref{S5Dcomb}) simply represents a comfortable way to connect the
PG and SG regions by a continuous line.

For the angular dependence of the transmission probability
$Tr(\varphi)$ we considered three scenarios:

\begin{figure*}[t]
    \centering
    \includegraphics[width=0.98\textwidth]{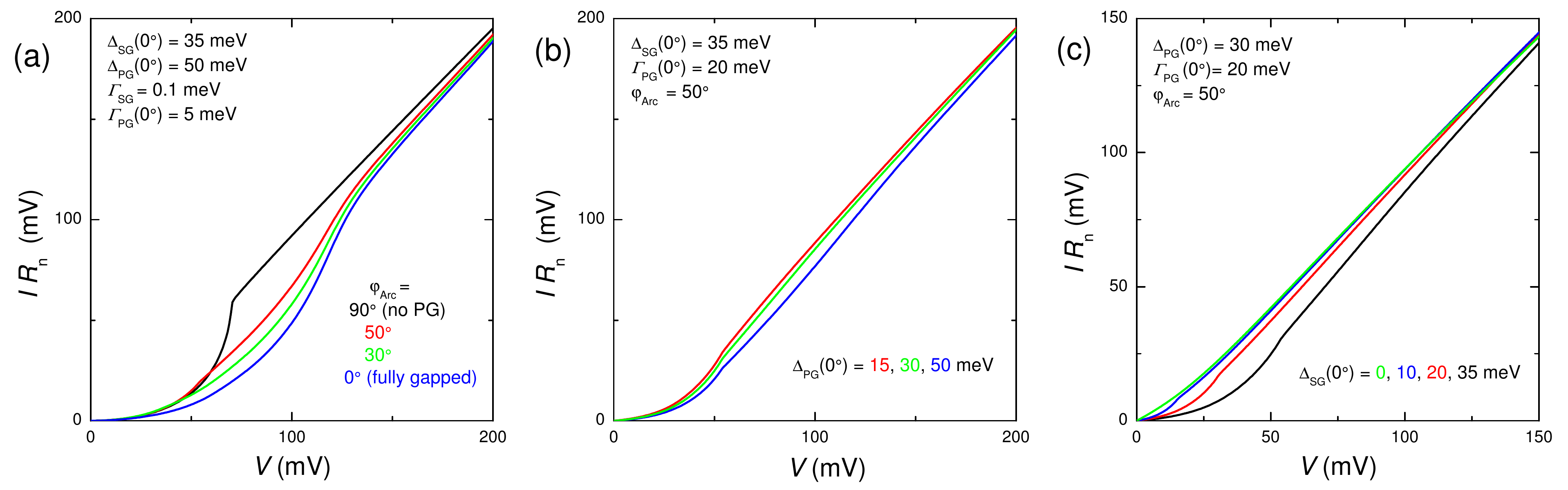}
    \caption{{\bf{Numerical modelling of intrinsic tunneling characteristics.}}
Evolution of calculated $I$-$V$ curves upon varying of {\bf(a)}
the pocket size, {\bf(b)} The pseudogap energy and {\bf(c)} the
superconducting gap. Note that the high-bias resistance remains
unchanged because it is integrated over the same barrel. }
    \label{fig:Sfig1}
\end{figure*}

(i) Non-directional tunneling $Tr(\varphi)=$const, as shown by the
dashed line in Fig. 2 (d).

(ii) Directional tunneling with maximum in antinodal and zero
transmission in nodal regions \cite{Ioffe_1998}, $Tr(\varphi)
\propto [\cos(k_x)-\cos(k_y)]^2 \propto
[\cos(\pi\sin(\varphi))-\cos(\pi\cos(\varphi))]^2$, as shown by
the dashed-dotted line in Fig. 2 (d). This expression is well
approximated by the function $\cos(2\varphi)^2$, which we will use
for analytical calculations below.

(iii) Semi-directional tunneling, which is an average of
non-directional and directional cases, $Tr(\varphi) \propto 1+
[\cos(\pi\sin(\varphi))-\cos(\pi\cos(\varphi))]^2/4$, as shown by
the solid line in Fig. 2 (d).

Finally, the QP density of states $N(\varphi)$ is likely to vary
along the barrel. However, in the absence of a confident knowledge
of the angular dependence of DoS we just assumed it to be constant
$N(\varphi)=N(0)$. The angular variation of DoS would have the
same affect as angular variation of the transmission coefficient.

Figure \ref{fig:Sfig1} shows calculated $I$-$V$ characteristics in
case of semi-directional tunneling, corresponding to $dI/dV$
curves in Fig. 2 (e-g). Figs. \ref{fig:Sfig1} (a-c) demonstrates
variation of $I$-$V$'s upon changing (a) the arc size, (b) the PG
energy and (c) $\Delta_{SG}$, while keeping other parameters
constant. Note that the high bias resistance remains the same,
irrespective of the PG. This occurs because in our case the PG is
a state-conserving gap, see Eq. (\ref{S2N(G)}). Therefore the same
current is recovered upon integration over the full barrel at high
enough bias, irrespective whether there is a PG or not. Note that
curves in Figs. \ref{fig:Sfig1} (b) and (c) calculated for a large
$\Gamma _{PG}(0)=20$ meV resemble the experimental characteristics
for UD mesas with a finite offset voltage $\Delta V_{PG}$, see
Figs. 1 (e) and (d).

\subsection{Dependence of the critical current on the arc size}

For a junction made of s-wave superconductors $I_c R_n$ is
independent of the transmission coefficient and is given by
Ambegaokar-Baratoff expression,
\begin{equation}\label{S6IcRnAB}
    I_c R_n (T \ll T_c) = \frac{\pi\Delta }{2e}.
\end{equation}

For coherent tunneling in a junction made of d-wave
superconductors, Eq. (\ref{S6IcRnAB}) is still valid for a
specific angle \cite{TanakaKashiwaya}, i.e.  for $\delta
I_c(\varphi) \delta R_n(\varphi)$, where $\delta I_c(\varphi)$ and
$\delta R_n(\varphi)$ are contributions to the critical current
and resistance from Cooper pairs and QP's with momentum in the
direction $\varphi$. Thus, $\delta I_c(\varphi) =\pi \Delta
(\varphi)/2e \delta R_n(\varphi)$. From Eq. (\ref{S1Eq.Tunn})
$1/\delta R_n (\varphi)=Ae N(0)^2 Tr(\varphi)^2~\delta
\varphi/2\pi $. Introducing a quantity $R_1=Ae N(0)^2$ for
resistance at unit transmission $Tr=1$, we can write expressions
for the total critical current and the normal resistance, taking
into account d-wave angular dependence of the gap $\Delta
(\varphi)=\Delta_0 \cos(2\varphi)$:

\begin{equation}\label{S7Ic_Int}
    I_c R_1=\frac{4}{\pi}\int_{\pi/4-\varphi_{Arc}/2}^{\pi/4}{Tr(\varphi)^2 \cos(2\varphi) d\varphi}
\end{equation}

\begin{equation}\label{S8Ic_Int}
    R_1 R_n^{-1}=\frac{4}{\pi}\int_0^{\pi/4}{Tr(\varphi)^2 d\varphi}
\end{equation}
Here we assumed that Cooper pair current is originating only from
the arc and single QP current is accumulated over the full length
of the barrel.

After integration of Eqs. (\ref{S7Ic_Int}) and (\ref{S8Ic_Int}) we
obtain :

\begin{equation}\label{S9IcRn_NDir}
I_c R_n=\frac{\Delta_0}{e}\left[1-\cos(\varphi_{Arc}) \right],
\end{equation}
for the nondirectional case $Tr(\varphi)\propto 1$. For
$\varphi_{Arc}=\pi/2$ it reduces to a known expression $I_c R_n
=\Delta_0/e $ \cite{TanakaKashiwaya}.

\begin{eqnarray*}\label{S10IcRn_Dir}
\nonumber I_c R_n = \frac{\Delta_0}{e}\frac{8}{3}*~~~~~~~~~~~~~~~~~~~~~~~~~~~~ \\
\left[\frac{8}{15}
-\cos(\varphi_{Arc})+\frac{2}{3}\cos(\varphi_{Arc})^3
-\frac{1}{5}\cos(\varphi_{Arc})^5 \right],
\end{eqnarray*}
for the directional case, $Tr(\varphi)\propto \cos(2\varphi)^2$,
and

\begin{eqnarray*}\label{S11IcRn_Dir}
\nonumber
I_c R_n = \frac{\Delta_0}{e}\frac{8}{19}* ~~~~~~~~~~~~~~~~~~~~~~~~~~~\\
\left[\frac{43}{15}
-4\cos(\varphi_{Arc})+\frac{4}{3}\cos(\varphi_{Arc})^3
-\frac{1}{5}\cos(\varphi_{Arc})^5 \right],
\end{eqnarray*}
for the semi-directional case, $Tr(\varphi)\propto
1+\cos(2\varphi)^2$. The corresponding curves $I_c R_n
(\varphi_{Arc})$ are shown in Fig. 3 (d). In all cases $I_c$
vanishes as arcs collapse $\varphi_{Arc}\rightarrow 0$ because we
assumed that the supercurrent is originating only from arc
regions. The $R_n$ is independent of $\varphi_{Arc}$ because the
QP current is integrated over the whole barrel, which we assumed
to be unchanged. As described in the manuscript, such model
provides a qualitative explanation of different doping
dependencies of the critical current density and the high-bias
differential resistance.




\end{document}